\documentclass[aps,prb,twocolumn,showpacs]{revtex4}

\usepackage{graphicx}
\usepackage{amsmath}
\usepackage{amssymb}

\begin{document}

\title{Characteristic energy of the nematic-order state and its connection to 
enhancement of superconductivity in cuprate superconductors}

\author{Zhangkai Cao$^{1}$, Xingyu Ma$^{1}$, Yiqun Liu$^{2}$, Huaiming Guo$^{3}$,
and Shiping Feng$^{1}$}
\email{spfeng@bnu.edu.cn}

\affiliation{$^{1}$Department of Physics, Beijing Normal University, Beijing 100875,
China}

\affiliation{$^{2}$School of Physics, Nanjing University, Nanjing 210093, China}

\affiliation{$^{3}$School of Physics, Beihang University, Beijing 100191, China}

\begin{abstract}
The new development in sublattice-phase-resolved imaging of electronic structure now allow
for the visualisation of the nematic-order state characteristic energy of cuprate
superconductors in a wide doping regime. However, it is still unclear how this
characteristic energy of the nematic-order state is correlated with the enhancement of
superconductivity. Here the doping dependence of the nematic-order state characteristic
energy in cuprate superconductors and of its possible connection to the enhancement of
superconductivity is investigated within the framework of the kinetic-energy-driven
superconductivity. It is shown that the characteristic energy of the nematic-order state
is found to be particularly large in the underdoped regime, then it smoothly decreases
upon the increase of doping, in full agreement with the corresponding experimental
observations. Moreover, the characteristic energy of the nematic-order state as a function
of the nematic-order state strength in the underdoped regime presents a similar behavior
of the superconducting transition temperature. This suggests a possible connection between
the nematic-order state characteristic energy and the enhancement of the superconductivity.
\end{abstract}


\maketitle


\section{Introduction}\label{Introduction}

In cuprate superconductors \cite{Bednorz86} , the strongly correlated motion of the electrons
is confined to the square-lattice CuO$_{2}$ planes \cite{Cooper94,Takenaka94}. However, this
strong electron correlation also induces the system to exhibit numerous ordering tendencies
\cite{Vishik18,Comin16,Kivelson19,Vojta09,Fradkin10,Fernandes19}. In addition to
superconductivity, a variety of spontaneous symmetry-breaking orders have been observed
experimentally, indicating the coexistence and intertwinement between these spontaneous
symmetry-breaking orders and superconductivity
\cite{Vishik18,Comin16,Kivelson19,Vojta09,Fradkin10,Fernandes19}. Among these spontaneous
symmetry-breaking orders, the most distinct form of order is electronic nematicity
\cite{Kivelson19,Vojta09,Fradkin10,Fernandes19}, which corresponds to that the electronic
structure preserves the translation symmetry but breaks the rotation symmetry of the
underlying square-lattice CuO$_{2}$ plane. This is why in the common practice, the
{\it strength of the electronic nematicity} is defined as the {\it orthorhomicity of the
electronic structure} \cite{Nakata18}. As a natural consequence of a doped Mott insulator,
the manipulation of the particular characteristics of the superconducting (SC) state with
coexisting electronically nematic order through the control of the doping and strength of
the electronic nematicity is hotly debated and has been believed to be key to the
understanding of the problem of why cuprate superconductors exhibit a number of the
anomalous properties \cite{Vishik18,Comin16,Kivelson19,Vojta09,Fradkin10,Fernandes19}.

Experimentally, the multiple measurement techniques have been used to elucidate the nature of
the quasiparticle excitation and of its interplay with spontaneous symmetry-breaking orders
and superconductivity
\cite{Nakata18,Hinkov08,Sato17,Daou10,Taillefer15,Wang21,Ando02,Wu17,Lawler10,Fujita14,Zheng17,Wu11,Comin14,Gerber15,Peng16,Caprara17,Arpaia19},
where it has been found that the electronically nematic order coexists
with the translation symmetry breaking such as charge order (or equivalently charge density
wave) in the well-defined regimes of the phase diagrams, appearing below the pseudogap
crossover temperature $T^{*}$ in the underdoped regime, and coexists with charge order and
superconductivity below the SC transition temperature $T_{\rm c}$. It thus shows that the
electronic nematicity is an integral part of the essential physics of cuprate superconductors.
The temperature scale for the onset of the dynamical charge order may increase monotonically
with the decrease of doping in the underdoped regime, however, the static charge order may
exhibit a dome-like shape temperature dependence
\cite{Wu11,Comin14,Gerber15,Peng16,Caprara17,Arpaia19}. In particular, this interplay of the
electronically nematic order and charge order in the normal-state has been invoked recently to
give a consistent explanation of the transport anisotropy \cite{Wahlberg21}. However, although
a number of consequences from the electronic nematicity together with the associated
fluctuation phenomena have been identified in the early experimental measurements
\cite{Nakata18,Hinkov08,Sato17,Daou10,Taillefer15,Wang21,Ando02,Wu17,Lawler10,Fujita14,Zheng17},
the evolution of the characteristic quantities of the electronic nematicity itself with doping
in the entire range of the SC dome remains puzzling. Fortunately, the instrumentation for
sublattice-phase-resolved imaging of electronic structure has improved dramatically in recent
years, allowing this experimental technique to visualize simultaneously the doping and
energy dependence of the quasiparticle scattering interference (QSI) in the SC-state with
coexisting symmetry-breaking ordered states \cite{Fujita19}. In this case, as a compensation
for the early scanning tunneling spectroscopy (STS) experimental studies
\cite{Lawler10,Fujita14,Zheng17}, this experimental technique has been used to detect the doping
and energy dependence of the tunneling conductance of Bi$_{2}$Sr$_{2}$CaCu$_{2}$O$_{8+\delta}$
over a large field of view, perform a Fourier transform, and analyze data from distinct regions
of momentum space \cite{Fujita19}. Moreover, to establish the link between the pseudogap
and electronic nematicity, the doping and energy dependence of the averaged density of states
$\rho(E)$ and the doping and energy dependence of the nematic-order spectrum,
\begin{eqnarray}\label{NEOP}
N^{\rm (Z)}(E)={\rm Re}Z({\bf Q}^{\rm (B)}_{y},E)-{\rm Re}Z({\bf Q}^{\rm (B)}_{x},E),
\end{eqnarray}
in momentum space from the reciprocal lattice vectors ${\bf Q}^{\rm (B)}_{x}=[2\pi,0]$ and
${\bf Q}^{\rm (B)}_{y}=[0,2\pi]$ have been measured \cite{Fujita19}, where
$Z({\bf r},E)\equiv g({\bf r},E)/g({\bf r},-E)$ is a ratio of differential tunneling conductances
at opposite bias, while $g({\bf r},E)=dI({\bf r},E=eV)/dV$ is differential tunneling conductance.
This nematic-order spectrum $N^{\rm (Z)}(E)$ can be also defined as the {\it order parameter of
the electronic nematicity} \cite{Zheng17}. The pseudogap extracted directly from the measured data
of the averaged density of states shows that the pseudogap smoothly decreases upon the increase of
doping \cite{Fujita19}. On the other hand, the measured data of the order parameter of the
electronic nematicity $N^{\rm (Z)}(E)$ show that $N^{\rm (Z)}(E)$ has a dome-like shape energy
dependence \cite{Fujita19}, with the maximal $N^{\rm (Z)}(E)$ appearance at an energy
$E^{\rm (N)}_{\rm max}$. This energy $E^{\rm (N)}_{\rm max}$ associated with the maximal
$N^{\rm (Z)}(E)$ is so-called the {\it nematic-order state characteristic energy}. More importantly,
the evolution of the nematic-order state characteristic energy with doping is identified, where
measured on the samples whose doping spans the pseudogap regime, the nematic-state characteristic
energy $E^{\rm (N)}_{\rm max}$ and pseudogap energy $\bar{\Delta}_{\rm PG}$ are, within the
experimental error, identical \cite{Fujita19}. These experimental results therefore identify the
electronically nematic order exists across the entire range of the SC dome. On the basis of these
experimental results, it has been argued that the pseudogap is a consequence of a tendency towards
an electronically ordered state that is a coexistence of the nematic order and charge orders and
breaks both the translation and rotation symmetry \cite{Kivelson19,Fujita19}.

Although the doping dependence of the nematic-order state characteristic energy in cuprate
superconductors has been well-identified experimentally in the entire range of the SC dome
\cite{Fujita19}, its full understanding is still a challenging issue. In particular, it is still
unclear how this nematic-order state characteristic energy $E^{\rm (N)}_{\rm max}$ evolves with
the strength of the electronic nematicity. Theoretically, the possible origins of the emergence
of the electronic nematicity have been suggested: the electronically nematic order occurs upon
melting of stripe order or charge order \cite{Kivelson98,Zaanen99,Kivelson03,Nie15}, or induces
by the electron Fermi surface (EFS) instability \cite{Halboth00,Kitatani17}, or is attributed to
the incommensurate pair-density-wave \cite{Dai18,Tu19}. In particular, it has been proposed that
the density of states near EFS and geometry of EFS (then the band structure) strongly affect the
nematic-order formation \cite{Auvray19,Bulut13}. This follows a basic fact that in the
square-lattice CuO$_{2}$ planes, both the density of states near EFS and geometry of EFS change
significantly when the quasiparticle dispersion possesses a saddle point induced by the van Hove
singularity. In our recent study \cite{Cao21}, the intertwinement of the electronic nematicity
with superconductivity in cuprate superconductors has been studied based on the
kinetic-energy-driven superconductivity, where we have shown that the electronic nematicity
enhances superconductivity. Moreover, we \cite{Cao21} have also shown that the order parameter of
the electronic nematicity achieves its maximum in the characteristic energy of the nematic-order
state, and then decreases rapidly as the energy moves away from the characteristic energy of the
nematic-order state, in agreement with the experimental observation \cite{Fujita19}. However, a
natural question is whether the characteristic energy of the nematic-order state is correlated
with the enhancement of superconductivity or not? In this paper, we study the doping dependence
of the nematic-order state characteristic energy and of its possible connection to the enhancement
of superconductivity along with this line, where one of our main results is that both the
characteristic energy of the nematic-order state and enhancement of $T_{\rm c}$ exhibit the same
nematic-order state strength dependence, i.e., the maximal characteristic energy of the
nematic-order state (then the maximal $T_{\rm c}$) occurs at around the optimal strength of the
electronic nematicity, and then decreases in both weak and strong strength regions. This suggests
a possible connection between the characteristic energy of the nematic-order state and the
enhancement of superconductivity.

This paper is organized as follows. We present the basic formalism in Sec. \ref{Formalism}, and
then discuss the doping dependence of the nematic-order state characteristic energy and of its
possible connection to the enhancement of superconductivity in Sec.
\ref{Quantitative-characteristics}, where we show that in a striking similar to the doping
dependence of the pseudogap state, the nematic-order state is particularly obvious in the
underdoped regime, i.e., the characteristic energy of the nematic-order state is particularly large
in the underdoped regime, and then it monotonically decreases with the increase of doping, in full
agreement with the corresponding experimental observations \cite{Fujita19}. Finally, we give a
summary and discussions in Sec. \ref{Conclusion}.

\section{Model and theoretical method}\label{Formalism}

When the quasiparticle scattering mixes the states ${\bf k}$ and ${\bf k}+{\bf q}$, a QSI pattern
with the wave vector ${\bf q}$ appears in the norm of the quasiparticle wave function and the
local density of states modulations with the wave length $\lambda=2\pi/|{\bf q}|$ appear,
reflecting a basic fact that the QSI pattern manifests itself is an autocorrelation between the
quasiparticle bands $E_{\bf k}$ and $E_{{\bf k}+{\bf q}}$ \cite{Pan01,Fischer07,Yin21}. In other
words, the intensity in the QSI pattern is proportional to the spectral intensities of the
single-particle excitation spectra at the momenta ${\bf k}$ and ${\bf k}+{\bf q}$, while the sharp
intensity peaks in the QSI pattern then are corresponding to the highest joint density of states.
This is why the quasiparticle scattering processes, the quasiparticle momentum-space structure,
and the dispersion of the peaks in the QSI pattern as a function of energy are interpreted in
terms of the octet scattering model \cite{Gao19,Wang03}, and yields the crucial information of the
quasiparticle excitation. More importantly, by the analysis of the typical feature of the Bragg
peaks in a QSI pattern, one is considering the phenomena that occur with the periodicity of the
underlying square lattice and which qualify any rotation symmetry-breaking
\cite{Pan01,Fischer07,Yin21}. Likewise, the autocorrelation between the quasiparticle bands
$E_{\bf k}$ and $E_{{\bf k}+{\bf q}}$ can be also measured in terms of the angle-resolved
photoemission spectroscopy (ARPES) autocorrelation \cite{Chatterjee06},
\begin{eqnarray}\label{ARPES-autocorrelation}
{\bar C}_{\varsigma}({\bf q},\omega)={1\over N}\sum_{\bf k}
I_{\varsigma}({\bf k}+{\bf q},\omega)I_{\varsigma}({\bf k},\omega),
\end{eqnarray}
where $N$ is the number of lattice sites, and $I_{\varsigma}({\bf k},\omega)$ is the
single-particle excitation spectrum, while the summation of momentum ${\bf k}$ is extended up
to the second Brillouin zone (BZ) \cite{Chatterjee06} for the discussion of QSI together with
the Bragg scattering. This ARPES autocorrelation in Eq. (\ref{ARPES-autocorrelation})
describes the correlation of the spectral intensities of the single-particle excitation spectra
at two different momenta ${\bf k}$ and ${\bf k}+{\bf q}$, separated by a momentum transfer
${\bf q}$, at a fixed energy, and is effectively the momentum-resolved joint density of states.
In particular, it has been demonstrated experimentally \cite{Chatterjee06,He14} that the peaks,
the momentum-space structure, and the dispersion of the peaks in the ARPES autocorrelation
pattern are directly related to the peaks, the momentum-space structure, and the dispersion of
the peaks in the QSI pattern \cite{Pan01,Fischer07,Yin21},
respectively, and can be also explained straightforwardly in terms of the octet scattering
model \cite{Gao19}. This is why the characteristic features of QSI can be also obtained in terms
of the ARPES autocorrelation.

The single-particle excitation spectrum $I_{\varsigma}({\bf k},\omega)$ in
Eq. (\ref{ARPES-autocorrelation}) is proportional to the electron spectral function
$A_{\varsigma}({\bf k},\omega)$ as,
\begin{eqnarray}\label{ARPES-spectrum}
I_{\varsigma}({\bf k},\omega)\propto n_{\rm F}(\omega)A_{\varsigma}({\bf k},\omega),
\end{eqnarray}
where $n_{\rm F}(\omega)$ is the fermion distribution, while the electron spectral function
$A_{\varsigma}({\bf k},\omega)$ in the SC-state with coexisting electronic nematicity can be
obtained directly from the full electron diagonal propagator as
$A_{\varsigma}({\bf k},\omega)=-2{\rm Im}G_{\varsigma}({\bf k},\omega)$.

Now our goal is to evaluate this full electron diagonal propagator
$G_{\varsigma}({\bf k},\omega)$ starting from a microscopic SC theory. The strongly correlated
motion of the electrons in cuprate superconductors is restricted to the square-lattice
CuO$_{2}$ planes \cite{Cooper94,Takenaka94} as mentioned above, and then the unconventional
properties come from the strongly correlated motion of the electrons in these CuO$_{2}$ planes.
In particular, as originally emphasized by Anderson \cite{Anderson87}, the essential physics of
the strongly correlated motion of the electrons in a square-lattice CuO$_{2}$ plane can be
described properly by the $t$-$J$ model,
\begin{eqnarray}\label{tJ-model}
H&=&-\sum_{l\hat{\eta}\sigma}t_{\hat{\eta}}C^{\dagger}_{l\sigma}C_{l+\hat{\eta}\sigma}
+\sum_{l\hat{\tau}\sigma}t'_{\hat{\tau}}C^{\dagger}_{l\sigma}C_{l+\hat{\tau}\sigma}
\nonumber\\
&+&\mu\sum_{l\sigma}C^{\dagger}_{l\sigma}C_{l\sigma} +\sum_{l\hat{\eta}}J_{\hat{\eta}}
{\bf S}_{l}\cdot {\bf S}_{l+\hat{\eta}},
\end{eqnarray}
where $\hat{\eta}=\pm\hat{x},\pm\hat{y}$ represents the nearest neighbor (NN) sites of a given
site $l$, $\hat{\tau}=\pm \hat{x}\pm\hat{y}$ represents the next NN sites of a given site $l$,
$C^{\dagger}_{l\sigma}$ and $C_{l\sigma}$ are the electron creation and annihilation operators,
respectively, ${\bf S}_{l}$ is the spin operator with its components $S^{\rm x}_{l}$,
$S^{\rm y}_{l}$, and $S^{\rm z}_{l}$, while $\mu$ is the chemical potential. For the
discussions of the exotic features of the SC-state with coexisting electronic nematicity, the
next NN hoping amplitude in the $t$-$J$ model (\ref{tJ-model}) is chosen as
$t'_{\hat{\tau}}=t'$, while the NN hoping amplitude $t_{\hat{\eta}}$ has the following form
\cite{Nakata18},
\begin{eqnarray}
t_{\hat{x}}&=&(1-\varsigma)t, ~~~~ t_{\hat{y}}=(1+\varsigma)t, \label{NN-hoping}
\end{eqnarray}
which is strongly anisotropic along the ${\hat{x}}$ and ${\hat{y}}$ directions and follows
from the previous analyses of the exotic features in the nematic-order state
\cite{Yamase00,Edegger06,Wollny09,Lee16}. In particular, this anisotropic NN hoping amplitude
in Eq. (\ref{NN-hoping}) has been experimentally verified in terms of the standard
tight-binding model to fit the ARPES spectrum in the nematic-order state \cite{Nakata18}.
Concomitantly, this anisotropic NN hoping amplitude in Eq. (\ref{NN-hoping}) induces the
anisotropic NN exchange coupling $J_{\hat{x}}=(1-\varsigma)^{2}J$ and
$J_{\hat{y}}=(1+\varsigma)^{2}J$ in the $t$-$J$ model (\ref{tJ-model}). Moreover, {\it this
anisotropic parameter} $\varsigma$ in Eq. (\ref{NN-hoping}) {\it represents the orthorhombicity
of the electronic structure}, {\it and therefore can be defined as the strength of the
electronic nematicity} in the system \cite{Nakata18}. In this sense, the anisotropic NN hoping
amplitudes in Eq. (\ref{NN-hoping}) also indicate that the rotation symmetry is broken already
in the starting $t$-$J$ model (\ref{tJ-model}). In cuprate superconductors, although the values
of $J$, $t$, and $t'$ are believed to vary somewhat from compound to compound, the commonly used
parameters in this paper are chosen as $t/J=3$, $t'/t=1/3$, and $J=100$ meV as in our previous
discussions \cite{Cao21}. Moreover, the temperature $T$ is set at $T=0.002J$. Unless otherwise
indicated, the doping is fixed at $\delta=0.06$ for a direct comparison with the corresponding
experimental result \cite{Fujita19}.

The $t$-$J$ model (\ref{tJ-model}) is supplemented by a on-site local constraint of no double
electron occupancy \cite{Feng93,Yu92,Lee06}, i.e.,
$\sum_{\sigma}C^{\dagger}_{l\sigma}C_{l\sigma}\leq 1$. However, the most difficult in the
analytical treatment of the $t$-$J$ model (\ref{tJ-model}) comes mainly from this local
constraint of no double electron occupancy, while the strong electron correlation manifests
itself by this local constraint of no double electron occupancy, and therefore the crucial
requirement is to impose this local constraint of no double electron occupancy. To incorporate
this local constraint of no double electron occupancy, the fermion-spin transformation
\cite{Feng0494,Feng15} has been proposed, where the physics of no double electron occupancy is
taken into account by representing the electron as a composite object created by,
\begin{eqnarray}\label{CSS}
C_{l\uparrow}=h^{\dagger}_{l\uparrow}S^{-}_{l}, ~~~~
C_{l\downarrow}=h^{\dagger}_{l\downarrow}S^{+}_{l},
\end{eqnarray}
with the spinful fermion operator $h_{l\sigma}=e^{-i\Phi_{l\sigma}}h_{l}$ that represents the
charge degree of freedom of the constrained electron together with some effects of spin
configuration rearrangements due to the presence of the doped hole itself (charge carrier), while
the spin operator $S_{l}$ describes the spin degree of freedom of the constrained electron, and
then the local constraint of no double occupancy is satisfied in analytical calculations. In this
fermion-spin representation (\ref{CSS}), the original $t$-$J$ model in Eq. (\ref{tJ-model}) can be
rewritten as,
\begin{eqnarray}\label{CSS-tJ-model}
H&=&\sum_{l\hat{\eta}}t_{\hat{\eta}}(h^{\dagger}_{l+\hat{\eta}\uparrow}h_{l\uparrow}S^{+}_{l}
S^{-}_{l+\hat{\eta}}+h^{\dagger}_{l+\hat{\eta}\downarrow}h_{l\downarrow}S^{-}_{l}
S^{+}_{l+\hat{\eta}})\nonumber\\
&-&\sum_{l\hat{\tau}}t'_{\hat{\tau}}(h^{\dagger}_{l+\hat{\tau}\uparrow}h_{l\uparrow}S^{+}_{l}
S^{-}_{l+\hat{\tau}}+ h^{\dagger}_{l+\hat{\tau}\downarrow}h_{l\downarrow}S^{-}_{l}
S^{+}_{l+\hat{\tau}})\nonumber\\
&-&\mu_{\rm h}\sum_{l\sigma}h^{\dagger}_{l\sigma}h_{l\sigma}+\sum_{l\hat{\eta}}
J^{(\hat{\eta})}_{\rm eff}{\bf S}_{l}\cdot {\bf S}_{l+\hat{\eta}},~~~~~~
\end{eqnarray}
where $\mu_{\rm h}$ is the charge-carrier chemical potential,
$S^{-}_{l}=S^{\rm x}_{l}-iS^{\rm y}_{l}$ and $S^{+}_{l}=S^{\rm x}_{l}+iS^{\rm y}_{l}$ are the
spin-lowering and spin-raising operators for the spin $S=1/2$, respectively,
$J_{\rm eff}^{(\hat{\eta})}=(1-\delta)^{2}J_{\hat{\eta}}$, and
$\delta=\langle h^{\dagger}_{l\sigma}h_{l\sigma}\rangle$ is the charge-carrier doping
concentration.

Within the $t$-$J$ model in the fermion-spin representation, the kinetic-energy-driven SC
mechanism has been developed in the case of the absence of the electronic nematicity
\cite{Feng15,Feng0306,Feng12,Feng15a}, where the interaction between the charge carriers
directly from the kinetic energy of the $t$-$J$ model by the exchange of a strongly dispersive
{\it spin excitation} generates the d-wave charge-carrier pairing in the particle-particle
channel, then the d-wave electron pairs originated from the d-wave charge-carrier pairing
state are due to the charge-spin recombination, and their condensation reveals the d-wave
SC-state. The typical features of the kinetic-energy-driven superconductivity can be also
summarized as: (i) the mechanism of the kinetic-energy-driven superconductivity is purely
electronic without phonons; (ii) the mechanism of the kinetic-energy-driven superconductivity
shows that the strong electron correlation is favorable to superconductivity, since the
bosonic glue is identified into an electron pairing mechanism not involving the phonon, the
external degree of freedom, but the internal spin degree of freedom of the constrained
electron; (iii) the SC-state is controlled by both the SC gap and quasiparticle coherence,
which leads to that the maximal $T_{\rm c}$ occurs around the optimal doping, and then
decreases in both the underdoped and the overdoped regimes. Very recently, the framework of
the kinetic-energy-driven superconductivity \cite{Feng15,Feng0306,Feng12,Feng15a} has been
generalized to discuss the intertwinement of the electronic nematicity with superconductivity
in cuprate superconductors \cite{Cao21}, where the breaking of the rotation symmetry due to
the presence of the electronic nematicity is verified by the inequivalence on the average of
the electronic structure at the two Bragg scattering sites. Our following discussions builds
on the work in Ref. \onlinecite{Cao21}, and only a short summary of the formalism is therefore
given. In the recent discussions \cite{Cao21}, the full electron diagonal and off-diagonal
propagators of the $t$-$J$ model (\ref{tJ-model}) have been given explicitly as,
\begin{subequations}\label{EGF}
\begin{eqnarray}
G_{\varsigma}({\bf k},\omega)&=&{1\over \omega-\varepsilon^{(\varsigma)}_{\bf k}
-\Sigma^{(\varsigma)}_{\rm tot}({\bf k},\omega)}, \label{DEGF}\\
\Im^{\dagger}_{\varsigma}({\bf k},\omega)&=&{L^{(\varsigma)}_{\bf k}(\omega)\over
\omega-\varepsilon^{(\varsigma)}_{\bf k}-\Sigma^{(\varsigma)}_{\rm tot}({\bf k},\omega)},
\label{ODEGF}
\end{eqnarray}
\end{subequations}
where the orthorhombic energy dispersion in the tight-binding approximation is obtained
directly from the $t$-$J$ model (\ref{tJ-model}) as,
\begin{eqnarray}
\varepsilon^{(\varsigma)}_{\bf k}=-4t[(1-\varsigma)\gamma_{{\bf k}_{x}}+(1+\varsigma)
\gamma_{{\bf k}_{y}}]+4t'\gamma_{\bf k}'+\mu,
\end{eqnarray}
with $\gamma_{{\bf k}_{x}}={\rm cos}k_{x}/2$, $\gamma_{{\bf k}_{y}}={\rm cos}k_{y}/2$,
$\gamma_{\bf k}'={\rm cos}k_{x}{\rm cos}k_{y}$, while the total self-energy
$\Sigma^{(\varsigma)}_{\rm tot}({\bf k},\omega)$ and weight function
$L^{(\varsigma)}_{\bf k}(\omega)$ are specific combinations of the normal self-energy
$\Sigma^{(\varsigma)}_{\rm ph}({\bf k},\omega)$ in the particle-hole channel and anomalous
self-energy $\Sigma^{(\varsigma)}_{\rm pp} ({\bf k},\omega)$ in the particle-particle channel
as,
\begin{subequations}
\begin{eqnarray}
\Sigma^{(\varsigma)}_{\rm tot}({\bf k},\omega)&=&\Sigma^{(\varsigma)}_{\rm ph}({\bf k},\omega)
+{|\Sigma^{(\varsigma)}_{\rm pp}({\bf k},\omega)|^{2}\over \omega
+\varepsilon^{(\varsigma)}_{\bf k}+\Sigma^{(\varsigma)}_{\rm ph}({\bf k}, -\omega)}, ~~~~~~
\label{TOT-SE}\\
L^{(\varsigma)}_{\bf k}(\omega)&=&-{\Sigma^{(\varsigma)}_{\rm pp}({\bf k},\omega)\over \omega
+\varepsilon^{(\varsigma)}_{\bf k}+\Sigma^{(\varsigma)}_{\rm ph}({\bf k},-\omega)},
\end{eqnarray}
\end{subequations}
where the normal self-energy $\Sigma^{(\varsigma)}_{\rm ph}({\bf k},\omega)$ and anomalous
self-energy $\Sigma^{(\varsigma)}_{\rm pp} ({\bf k},\omega)$ have been obtained in
Ref. \onlinecite{Cao21}, and can be expressed explicitly as,
\begin{widetext}
\begin{subequations}\label{ESE}
\begin{eqnarray}
{\Sigma}^{(\varsigma)}_{\rm ph}({\bf{k}},{\omega})&=&\frac{1}{2N^{2}}\sum_{{\bf{p}}{\bf{p}'}{\nu}}(-1)^{\nu+1}
{\Omega}^{(\varsigma)}_{{\bf{p}}{\bf{p}'}{\bf{k}}}\left [\left ( 1+{\bar{\varepsilon}^{(\varsigma)}_{\bf{p}+\bf{k}}
\over E^{(\varsigma)}_{\bf{p}+\bf{k}}}\right )
\left (\frac{F^{(\varsigma)}_{1\nu}({\bf p},{\bf p}',{\bf k})}{\omega+\omega^{(\nu)}_{\varsigma{\bf{p}}{\bf{p}}'}
-E^{(\varsigma)}_{\bf{p}+\bf{k}}}-\frac{F^{(\varsigma)}_{2\nu}({\bf p},{\bf p}',{\bf k})}
{\omega-\omega^{(\nu)}_{\varsigma{\bf{p}}{\bf{p}}'}-E^{(\varsigma)}_{\bf{p}+\bf{k}}}\right)\right.\nonumber\\
&+&\left. \left ( 1-{\bar{\varepsilon}^{(\varsigma)}_{\bf{p}+\bf{k}}\over E^{(\varsigma)}_{\bf{p}+\bf{k}}}\right )
\left (\frac{F^{(\varsigma)}_{1\nu}({\bf p},{\bf p}',{\bf k})}
{\omega- \omega^{(\nu)}_{\varsigma{\bf{p}}{\bf{p}}'}+E^{(\varsigma)}_{\bf{p}+\bf{k}}}
-\frac{F^{(\varsigma)}_{2\nu}({\bf p},{\bf p}',{\bf k})}{\omega+\omega^{(\nu)}_{\varsigma{\bf{p}}{\bf{p}}'}
+E^{(\varsigma)}_{\bf{p}+\bf{k}}}  \right )  \right ],\label{ph-ESE}\\
{\Sigma}^{(\varsigma)}_{\rm pp}({\bf{k}},{\omega})&=&\frac{1}{2N^{2}}\sum_{{\bf{p}}{\bf{p}'}{\nu}}(-1)^{\nu}
{\Omega}^{(\varsigma)}_{{\bf{p}}{\bf{p}'}{\bf{k}}}\frac{\bar{\Delta}_{\varsigma{\rm{Z}}}({\bf{p}}+{\bf{k}})}
{E^{(\varsigma)}_{{\bf{p}}+\bf{k}}}\left [ \left ( \frac{F^{(\varsigma)}_{1\nu}({\bf p},{\bf p}',{\bf k})} {\omega+\omega^{(\nu)}_{\varsigma{\bf{p}}{\bf{p}}'}-E^{(\varsigma)}_{\bf{p}+\bf{k}}}
-\frac{F^{(\varsigma)}_{2\nu}({\bf p},{\bf p}',{\bf k})}{\omega-\omega^{(\nu)}_{\varsigma{\bf{p}}{\bf{p}}'}
-E^{(\varsigma)}_{\bf{p}+\bf{k}}} \right)\right.\nonumber\\
&-& \left. \left ( \frac{F^{(\varsigma)}_{1\nu}({\bf p},{\bf p}',{\bf k})}{\omega-\omega^{(\nu)}_{\varsigma{\bf{p}}{\bf{p}}'} +E^{(\varsigma)}_{\bf{p}+\bf{k}}}- \frac{F^{(\varsigma)}_{2\nu}({\bf p},{\bf p}',{\bf{k}})}
{\omega+\omega^{(\nu)}_{\varsigma{\bf{p}}{\bf{p}}'}
+E^{(\varsigma)}_{\bf{p}+\bf{k}}}  \right )  \right ], \label{pp-ESE}
\end{eqnarray}
\end{subequations}
\end{widetext}
where $\nu=1,2$, ${\Omega}^{(\varsigma)}_{{\bf{p}}{\bf{p}'}{\bf{k}}}=Z^{(\varsigma)}_{\rm{F}}
[\Lambda^{(\varsigma)}_{{\bf{p}}+{\bf{p}'}+{\bf{k}}}]^{2}B^{(\varsigma)}_{\bf{p}'}
B^{(\varsigma)}_{\bf{p}+\bf{p}'}/(4\omega^{(\varsigma)}_{\bf{p}'}
\omega^{(\varsigma)}_{\bf{p}+\bf{p}'})$ with $\Lambda^{(\varsigma)}_{\bf k}=4t[(1-\varsigma)
\gamma_{{\bf k}_{x}}+(1+\varsigma)\gamma_{{\bf k}_{y}}]-4t'\gamma_{\bf{k}}'$,
the SC quasiparticle energy spectrum
$E^{(\varsigma)}_{\bf k}=\sqrt {\bar{\varepsilon}^{(\varsigma)2}_{\bf k}
+|\bar{\Delta}^{(\varsigma)}_{\rm Z}({\bf k})|^{2}}$ with the renormalized SC gap
$\bar{\Delta}^{(\varsigma)}_{\rm Z}({\bf k})=Z^{(\varsigma)}_{\rm F}\bar{\Delta}^{(\varsigma)}({\bf k})$
and renormalized electron orthorhombic energy dispersion
$\bar{\varepsilon}^{(\varsigma)}_{\bf k}=Z^{(\varsigma)}_{\rm F}\varepsilon^{(\varsigma)}_{\bf k}$,
$\omega^{(\nu)}_{\varsigma{\bf p}{\bf p}'} ={\omega}^{(\varsigma)}_{\bf{p}+\bf{p}'}
-(-1)^{\nu}{\omega}^{(\varsigma)}_{\bf{p}'}$, while the quasiparticle coherent weight
$Z^{(\varsigma)}_{\rm F}$, the SC gap $\bar{\Delta}^{(\varsigma)}({\bf k})$, the spin
orthorhombic excitation spectrum $\omega^{(\varsigma)}_{\bf k}$, the weight function of the spin
excitation spectrum $B^{(\varsigma)}_{{\bf{k}}}$, and the functions
$F^{(\varsigma)}_{1\nu}({\bf{p}},{\bf{p}}',{\bf{k}})$ and
$F^{(\varsigma)}_{2\nu}({\bf{p}},{\bf{p}}',{\bf{k}})$ have been given explicitly in
Ref. \onlinecite{Cao21}. In particular, the sharp peak visible for temperature $T\rightarrow 0$ in
the normal (anomalous) self-energy is actually a $\delta$-function, broadened by a small damping
used in the numerical calculation at a finite lattice. The calculation in this paper for the
normal (anomalous) self-energy is performed numerically on a $120\times 120$ lattice in momentum
space, with the infinitesimal $i0_{+}\rightarrow i\Gamma$ replaced by a small damping
$\Gamma=0.05J$.

With the above full electron diagonal Green's function (\ref{DEGF}), the electron spectral
function $A_{\varsigma}({\bf k},\omega)$ in the SC-state with coexisting electronic nematicity
now can be obtained explicitly as,
\begin{eqnarray}\label{ESF}
A_{\varsigma}({\bf k},\omega)={-2{\rm Im}\Sigma^{(\varsigma)}_{\rm tot}({\bf k},\omega)\over
[\omega-\varepsilon^{(\varsigma)}_{\bf k}-{\rm Re}
\Sigma^{(\varsigma)}_{\rm tot}({\bf k},\omega)]^{2}+[{\rm Im}
\Sigma^{(\varsigma)}_{\rm tot}({\bf k},\omega)]^{2}},~~
\end{eqnarray}
where ${\rm Re}\Sigma^{(\varsigma)}_{\rm tot}({\bf k},\omega)$ and
${\rm Im}\Sigma^{(\varsigma)}_{\rm tot}({\bf k},\omega)$ are the real and imaginary parts of the
total self-energy $\Sigma_{\rm tot}({\bf k},\omega)$, respectively. Substituting this electron
spectral function $A_{\varsigma}({\bf k},\omega)$ in Eq. (\ref{ESF}) into
Eqs. (\ref{ARPES-spectrum}) and (\ref{ARPES-autocorrelation}), we therefore obtain the ARPES
autocorrelation ${\bar C}_{\varsigma}({\bf q},\omega)$ within the framework of the
kinetic-energy-driven superconductivity.

\section{Quantitative characteristics}\label{Quantitative-characteristics}

\begin{figure}[h!]
\centering
\includegraphics[scale=0.90]{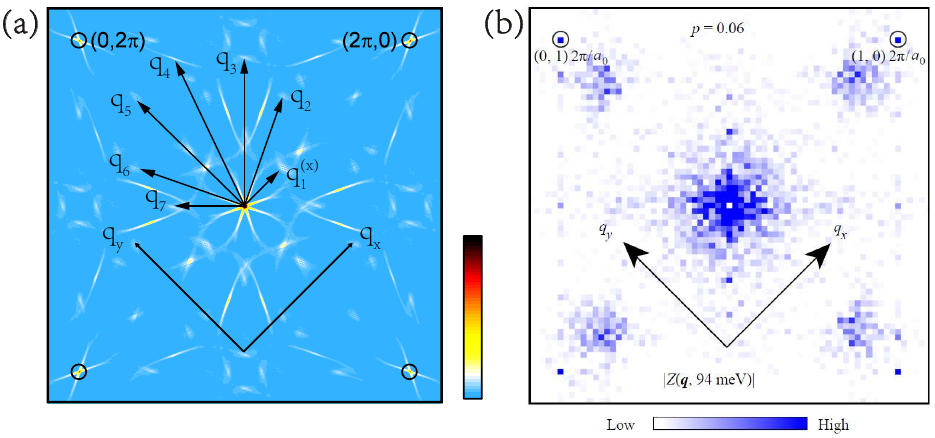}
\caption{(Color online) (a) The ARPES autocorrelation pattern in momentum-space in the
binding-energy $\omega=94$ meV for the strength of the electronic nematicity $\varsigma=0.006$.
(b) The corresponding experimental result of the quasiparticle scattering interference pattern
of Bi$_{2}$Sr$_{2}$CaCu$_{2}$O$_{8+\delta}$ in the binding-energy $\omega=94$ meV at doping
$\delta=0.06$ taken from Ref. \onlinecite{Fujita19}.
\label{autocorrelation-maps}}
\end{figure}

In the presence of the electronic nematicity, the original electronic structure with the
four-fold ($C_{4}$) rotation symmetry on the square lattice in the absence of the electronic
nematicity is broken up into that with a residual two-fold ($C_{2}$) rotation symmetry, while
such an aspect should be reflected in QSI. For convenience, we plot the ARPES autocorrelation
${\bar C}_{\varsigma}({\bf q},\omega)$ in Fig. \ref{autocorrelation-maps}a. We are
considering the binding-energy $\omega=94$ meV and the strength of the electronic nematicity
$\varsigma=0.006$. In Fig. \ref{autocorrelation-maps}a, the locations of the Bragg peaks
${\bf Q}^{\rm (B)}_{x}=[\pm 2\pi,0]$ along the $\hat{x}$ axis and
${\bf Q}^{\rm (B)}_{y}=[0,\pm 2\pi]$ along the $\hat{y}$ axis are indicated by the circles,
while ${\bf q}_{1}$, ${\bf q}_{2}$, ${\bf q}_{3}$, ${\bf q}_{4}$, ${\bf q}_{5}$, ${\bf q}_{6}$,
and ${\bf q}_{7}$ are different quasiparticle scattering wave vectors. For a better comparison,
the corresponding experimental result \cite{Fujita19} of the QSI pattern observed on
Bi$_{2}$Sr$_{2}$CaCu$_{2}$O$_{8+\delta}$ in the bind-energy $\omega=94$ meV at doping
$\delta=0.06$ is also shown in Fig. \ref{autocorrelation-maps}b. The results in
Fig. \ref{autocorrelation-maps} thus show that the momentum-space structure of the ARPES
autocorrelation pattern in the SC-state with coexisting electronic nematicity is qualitative
consistent with the corresponding momentum-space structure of the QSI pattern observed on
Bi$_{2}$Sr$_{2}$CaCu$_{2}$O$_{8+\delta}$. Moreover, the characteristic features of two distinct
classes of the broken-symmetry states have been summarized as \cite{Cao21}: (i) For the
quasiparticle scattering processes with the corresponding scattering wave vectors ${\bf q}_{1}$,
${\bf q}_{4}$, and ${\bf q}_{5}$, the amplitudes of the quasiparticle scattering wave vectors
are respectively inequivalent to their symmetry-corresponding partners, while for the
quasiparticle scattering process with the corresponding quasiparticle scattering wave vectors
${\bf q}_{2}$, ${\bf q}_{3}$, ${\bf q}_{6}$, and ${\bf q}_{7}$, the scattering wave vectors and
their symmetry-equivalent partners occur with equal amplitudes. These results therefore indicate
that the peaks at the corresponding scattering wave vectors ${\bf q}_{1}$, ${\bf q}_{4}$, and
${\bf q}_{5}$ are the signatures of the electronically ordered states with broken both rotation
and translation symmetries, while the peaks with the corresponding scattering wave vectors
${\bf q}_{2}$, ${\bf q}_{3}$, ${\bf q}_{6}$, and ${\bf q}_{7}$ are the signatures of the
electronically ordered states with broken translation symmetry only; (ii) The intensity of the
peak at the Bragg wave vector ${\bf Q}^{\rm (B)}_{x}$ is different from that at the Bragg wave
vector ${\bf Q}^{\rm (B)}_{y}$. This difference leads to the inequivalence on the average of the
electronic structure at the two Bragg scattering sites ${\bf Q}^{\rm (B)}_{x}$ and
${\bf Q}^{\rm (B)}_{y}$, and therefore shows that the Bragg peaks at the wave vectors
${\bf Q}^{\rm (B)}_{x}$ and ${\bf Q}^{\rm (B)}_{y}$ are the signature of the nematic-order state
with the broken $C_{4}$ rotation symmetry.

\begin{figure}[h!]
\centering
\includegraphics[scale=0.80]{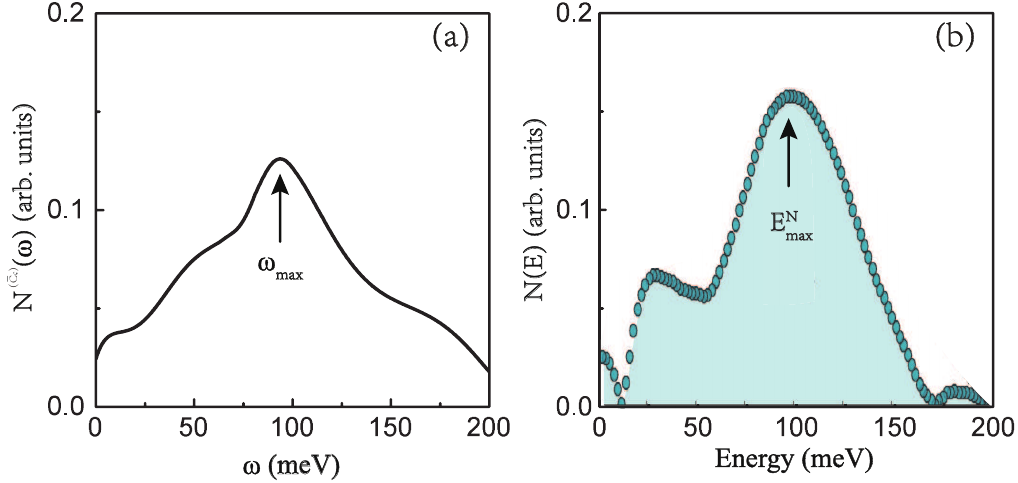}
\caption{(Color online) (a) The order parameter of the nematic-order state as a function of
energy for the strength of the electronic nematicity $\varsigma=0.006$, where the arrow
indicates the position of the peak at the characteristic energy. (b) The corresponding
experimental result observed on Bi$_{2}$Sr$_{2}$CaCu$_{2}$O$_{8+\delta}$ at doping
$\delta=0.06$ taken from Ref. \onlinecite{Fujita19}. \label{order-parameter}}
\end{figure}

We are now ready to discuss the doping dependence of the nematic-order state characteristic
energy and of its possible connection to the enhancement of superconductivity. In our previous
studies \cite{Cao21}, the order parameter of the electronic nematicity has been given as,
\begin{eqnarray}\label{nematic-order-parameter}
N^{({\bar C}_{\varsigma})}(\omega)={{\bar C}^{(x)}_{\varsigma}(\omega)
-{\bar C}^{(y)}_{\varsigma}(\omega)\over
{\bar C}^{(x)}_{\varsigma}(\omega)+{\bar C}^{(y)}_{\varsigma}(\omega)},
\end{eqnarray}
where ${\bar C}^{(x)}_{\varsigma}(\omega)=(1/N)\sum_{{\bf q}\in \{{\bf Q}^{\rm (B)}_{x}\}}
{\bar C}_{\varsigma}({\bf q},\omega)$ and ${\bar C}^{(y)}_{\varsigma}(\omega)=(1/N)
\sum_{{\bf q}\in \{{\bf Q}^{\rm (B)}_{y}\}}{\bar C}_{\varsigma}({\bf q},\omega)$, with the
summation ${\bf q}\in \{{\bf Q}^{\rm (B)}_{x}\}$ $[{\bf q}\in \{{\bf Q}^{\rm (B)}_{y}\}]$ that
is restricted to the extremely small area $\{{\bf Q}^{\rm (B)}_{x}\}$
$[\{{\bf Q}^{\rm (B)}_{y}\}]$ at around ${\bf Q}^{\rm (B)}_{x}$ $[{\bf Q}^{\rm (B)}_{y}]$. This
definition in Eq. (\ref{nematic-order-parameter}) is confronted with the reduction of the size
effect in a finite-lattice calculation. This follows a basic fact that the calculation for the
normal and anomalous self-energies in Eq. (\ref{ESE}) is performed numerically on a
$120\times 120$ lattice in momentum space as we have mentioned above, with the infinitesimal
$i0_{+}\rightarrow i\Gamma$ replaced by a small damping $\Gamma=0.05J$, which leads to that the
peak weight of the ARPES autocorrelation ${\bar C}_{\varsigma}({\bf q},\omega)$ in
Eq. (\ref{ARPES-autocorrelation}) at the Bragg wave vector ${\bf Q}^{\rm (B)}_{x}$
$[{\bf Q}^{\rm (B)}_{y}]$ spreads on the extremely small area $\{{\bf Q}^{\rm (B)}_{x}\}$
$[\{{\bf Q}^{\rm (B)}_{y}\}]$ at around the ${\bf Q}^{\rm (B)}_{x}$ $[{\bf Q}^{\rm (B)}_{y}]$
point. The summation of these spread weights in ${\bar C}^{(x)}_{\varsigma}(\omega)$
$[{\bar C}^{(y)}_{\varsigma}(\omega)]$ at around this extremely small area
$\{{\bf Q}^{\rm (B)}_{x}\}$ $[\{{\bf Q}^{\rm (B)}_{y}\}]$ can reduce the size effect in the
finite-lattice calculation. If this order parameter $N^{({\bar C}_{\varsigma})}(\omega)$ is
non-zero, the break of the $C_{4}$ rotation symmetry is occurring. In Fig. \ref{order-parameter}a,
we plot the order parameter of the nematic-order state $N^{({\bar C}_{\varsigma})}(\omega)$ as a
function of binding-energy for the strength of the electronic nematicity $\varsigma=0.006$.
For a direct comparison, the corresponding experimental result
\cite{Fujita19} of the energy dependence of the nematic-order state order parameter observed
on Bi$_{2}$Sr$_{2}$CaCu$_{2}$O$_{8+\delta}$ at doping $\delta=0.06$ is also shown in
Fig. \ref{order-parameter}b. It thus shows clearly that the experimental result \cite{Fujita19}
of the energy dependence of the nematic-order state order parameter is well reproduced, where
$N^{({\bar C}_{\varsigma})}(\omega)$ reaches its maximum in the characteristic energy
$\omega_{\rm max}$, however, when the energy is turned away from this characteristic energy
$\omega_{\rm max}$, $N^{({\bar C}_{\varsigma})}(\omega)$ drops rapidly. Moreover, this anticipated
characteristic energy $\omega_{\rm max}\sim 0.936J=93.6$ meV is well consistent with the
experimental result \cite{Fujita19} of $E^{(N)}_{\rm max}\sim 94$ meV observed on
Bi$_{2}$Sr$_{2}$CaCu$_{2}$O$_{8+\delta}$ at doping $\delta=0.06$. This energy dependence of
$N^{({\bar C}_{\varsigma})}(\omega)$ with non-zero values therefore further verifies the
nematic-order state with broken $C_{4}$ rotation symmetry in a wide energy range.

\begin{figure}[h!]
\centering
\includegraphics[scale=0.67]{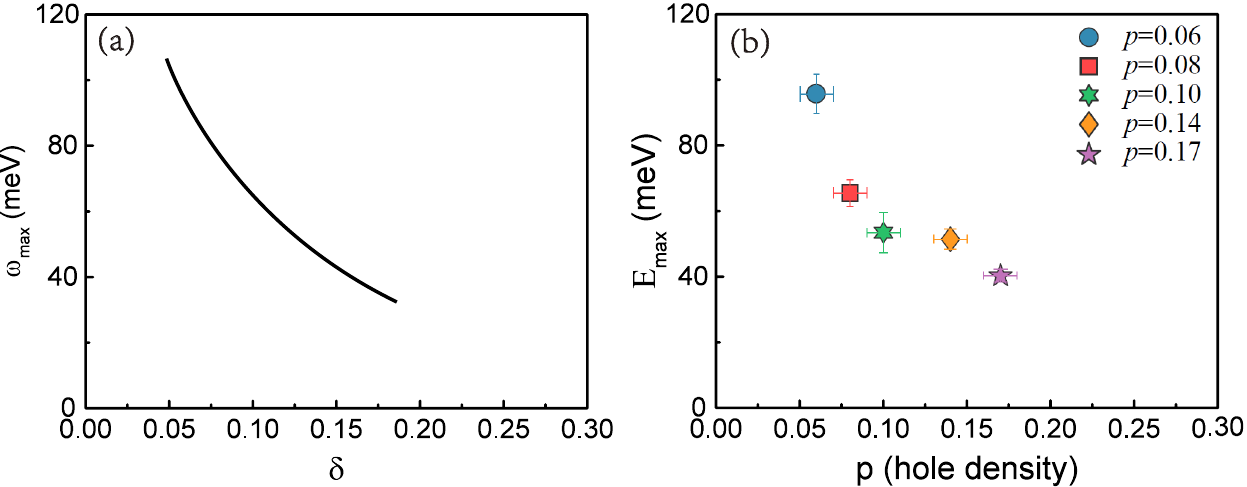}
\caption{(Color online) (a) The characteristic energy of the nematic-order state as a function
of doping for the strength of the electronic nematicity $\varsigma=0.006$. (b) The corresponding
experimental result observed on Bi$_{2}$Sr$_{2}$CaCu$_{2}$O$_{8+\delta}$ taken from
Ref. \onlinecite{Fujita19}. \label{energy-doping}}
\end{figure}

As a natural consequence of the doped Mott insulators, the characteristic energy of the
nematic-order state $\omega_{\rm max}$ also evolve strongly with doping. For a better
understanding of the doping dependence of $\omega_{\rm max}$, we plot the result of
$\omega_{\rm max}$ as a function of doping for the strength of the electronic nematicity
$\varsigma=0.006$ in Fig. \ref{energy-doping}a in comparison with the corresponding experimental
result \cite{Fujita19} of the doping dependence of the nematic-order state characteristic energy
observed on Bi$_{2}$Sr$_{2}$CaCu$_{2}$O$_{8+\delta}$ in Fig. \ref{energy-doping}b. The result
in Fig. \ref{energy-doping}a indicates clearly that $\omega_{\rm max}$ is particularly large in
the underdoped regime, and then monotonically decreases as doping is increased, which is fully
consistent with the corresponding result observed on Bi$_{2}$Sr$_{2}$CaCu$_{2}$O$_{8+\delta}$.
Moreover, we have also compared the above result of the doping dependence of the nematic-order
state characteristic energy with the experimental result of the doping dependence of the
pseudogap \cite{Fujita19}, and found that the nematic-order state characteristic energy and
pseudogap energy are also identical. The pseudogap in the framework of the kinetic-energy-driven
superconductivity originates from the electron self-energy resulting of the dressing of the
electrons due to the electron interaction mediated by a strongly dispersive spin excitation
\cite{Feng15,Feng0306,Feng12,Feng15a}, and then it can be identified as being a region of the
electron self-energy effect \cite{Timusk99,Hufner08} in which the pseudogap suppresses strongly
the electronic density of states. The characteristic energy of the nematic-order state in theory
and experiment is virtually identical to each other and also to the corresponding pseudogap
energy. These results therefore are important to confirm the nematic-order state characteristic
energy at the $C_{4}$ rotation symmetry-breaking can be understood as the natural consequence
of the electronic nematic-order state within the pseudogap of cuprate superconductors
\cite{Fujita19}.

In our recent studies \cite{Cao21}, the evolution of $T_{\rm c}$ with the strength of the
electronic nematicity has been obtained within the framework of the kinetic-energy-driven
superconductivity in terms of the self-consistent calculation at the condition of the SC gap
$\bar{\Delta}^{(\varsigma)}=0$, where the optimized $T_{\rm c}$ at the optimal doping
$\delta\approx 0.15$ increases with the increase of the strength of the electronic nematicity, and
reaches its maximum in the optimal strength of the electronic nematicity $\varsigma\approx 0.022$,
subsequently, the optimized $T_{\rm c}$ decreases with the increase of the strength of the
electronic nematicity in the strong strength region. This dome-like shape nematic-order strength
dependence of $T_{\rm c}$ therefore shows clearly that superconductivity in cuprate superconductors
is enhanced by the electronic nematicity. In particular, it has been shown that the energy in the
SC-state with coexisting electronic nematicity is lower than the corresponding energy in the
SC-state with the absence of the electronic nematicity \cite{Cao21}. Moreover, the SC condensation
energy as a function of the nematic-order state strength shows the same behavior of $T_{\rm c}$.
This same dome-like shape nematic-order strength dependence of the SC condensation energy thus
leads to that superconductivity is enhanced by the electronic nematicity, and $T_{\rm c}$ exhibits
a dome-like shape nematic-order strength dependence.

\begin{figure}[h!]
\centering
\includegraphics[scale=0.67]{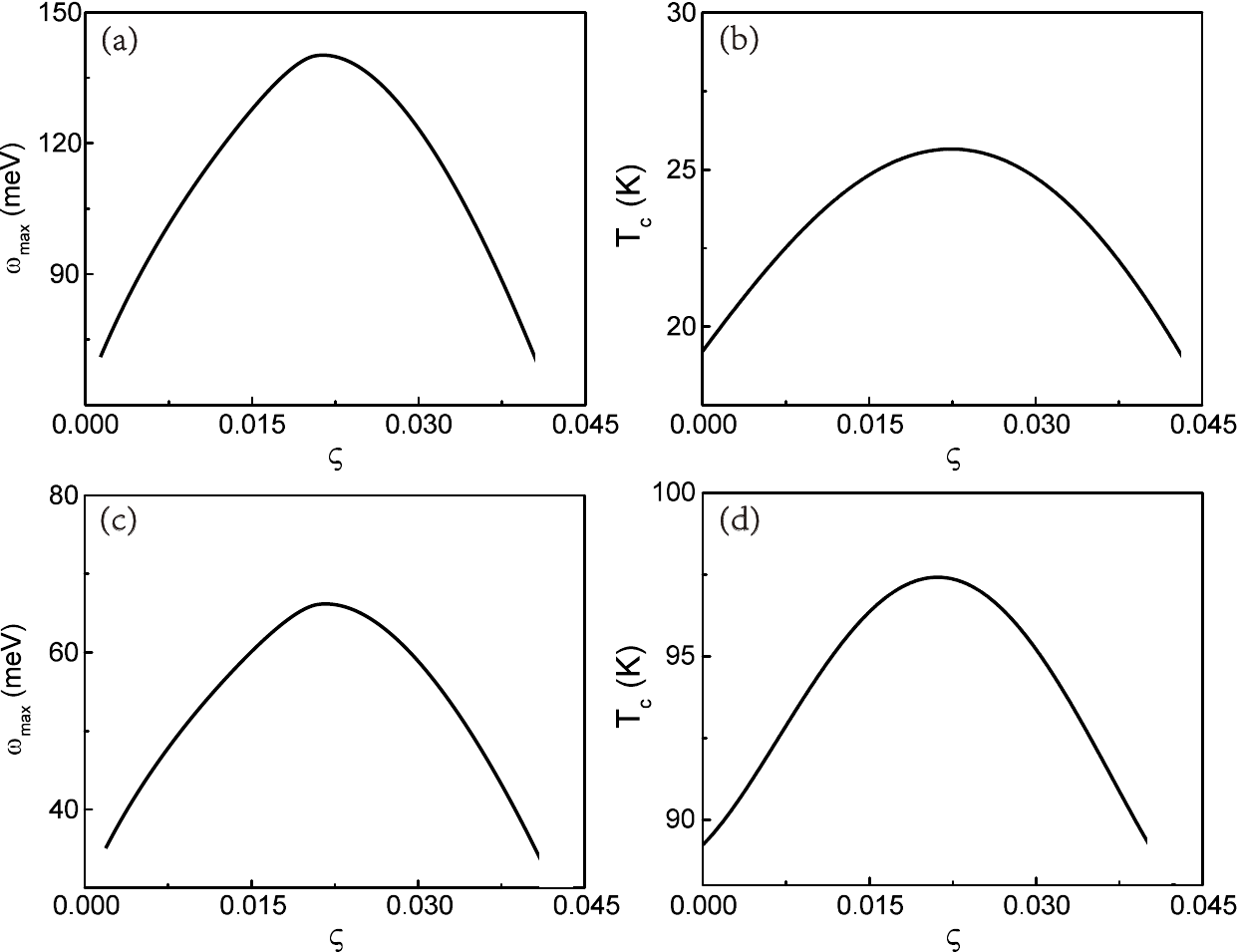}
\caption{(Color online) (a) The nematic-order state characteristic energy as a function of the
strength of the electronic nematicity $\varsigma$ at the underdoping $\delta=0.06$. (b) The
superconducting transition temperature as a function of the strength of the electronic nematicity
at the underdoping $\delta=0.06$. The corresponding results of (c) the nematic-order state
characteristic energy as a function of the strength of the electronic nematicity $\varsigma$ at
the optimal doping $\delta=0.15$ and (d) the superconducting transition temperature as a function
of the strength of the electronic nematicity at the optimal doping $\delta=0.15$ taken from
Ref. \onlinecite{Cao21}.  \label{energy-strength}}
\end{figure}

Now we turn our attention to the possible connection between the nematic-order state characteristic
energy and the enhancement of superconductivity. To show this possible connection more clearly, we
plot (a) $\omega_{\rm max}$ and (b) $T_{\rm c}$ as a function of the strength of the electronic
nematicity $\varsigma$ at the {\it underdoping} $\delta=0.06$ in Fig. \ref{energy-strength}. In
order to compare clearly the present results of the nematic-order state strength dependence of
$\omega_{\rm max}$ and $T_{\rm c}$ with the corresponding results at different doping levels,
the previous results \cite{Cao21} of (c) $\omega_{\rm max}$ and (d) $T_{\rm c}$ as a function of
the strength of the electronic nematicity at the {\it optimal doping} $\delta=0.15$ are also shown
in Fig. \ref{energy-strength}. Obviously, two characteristic features in Fig. \ref{energy-strength}
can be summarized as: (i) for the present case at the underdoping $\delta=0.06$ (see
Fig. \ref{energy-strength}a and Fig. \ref{energy-strength}b), the strength range together with the
{\it optimal strength} in $\omega_{\rm max}$ are the same with that in $T_{\rm c}$. In particular,
with the increase of the nematic-order state strength, $\omega_{\rm max}$ (then $T_{\rm c}$) is
raised gradually in the weak strength region, and achieves its maximum at around the
{\it optimal strength} $\varsigma=0.022$. However, with the further increase of the strength,
$\omega_{\rm max}$ (then $T_{\rm c}$) turns into a monotonically decrease in the strong strength
region; (ii) In comparison with the results \cite{Cao21} at the optimal doping $\delta=0.15$
(see Fig. \ref{energy-strength}c and Fig. \ref{energy-strength}d), $\omega_{\rm max}$ in
Fig. \ref{energy-strength}a ($T_{\rm c}$ in Fig. \ref{energy-strength}b) for a given nematic-order
state strength at the underdoping $\delta=0.06$ is much larger (lower) than the corresponding
$\omega_{\rm max}$ in Fig. \ref{energy-strength}c ($T_{\rm c}$ in Fig. \ref{energy-strength}d) at
the optimal doping $\delta=0.15$. However, the global dome-like shape of the nematic-order strength
dependence of $\omega_{\rm max}$ and $T_{\rm c}$ at the underdoping $\delta=0.06$ together with the
magnitude of the {\it optimal} strength are the same with that at the optimal doping $\delta=0.15$.
Therefore the enhancement of superconductivity occurs at a any given doping of the SC dome. This
same strength range together with the same {\it optimal} strength in the characteristic energy
$\omega_{\rm max}$ and SC transition temperature $T_{\rm c}$ therefore indicates firstly a possible
connection between the nematic-order state characteristic energy and the enhancement of
superconductivity.

\section{Summary and discussions}\label{Conclusion}

Within the framework of the kinetic-energy-driven superconductivity, we have studied
the doping dependence of the nematic-order state characteristic energy in cuprate superconductors
and of its possible connection to the enhancement of superconductivity. Our results show clearly
that the characteristic energy of the nematic-order state is particularly large in the underdoped
regime, then it smoothly decreases as doping is increased, in full agreement with the corresponding
to the STS experimental observations. More importantly, our results also indicate firstly that the
characteristic energy of the nematic-order state as a function of the nematic-order state strength
in the underdoped regime presents a similar behavior of the SC transition temperature. On the basis
of these obtained results, the theory therefore predicts a possible connection between the
nematic-order state characteristic energy and the enhancement of superconductivity.

Finally, it should be noted that apart from the emergence of the electronically nematic order in
cuprate superconductors \cite{Vojta09,Fradkin10,Fernandes19}, the electronic nematicity has been
observed across several families of strongly correlated electron systems, including iron-based
superconductors \cite{Chuang10,Gallais13,Massat16}, strontium ruthenates \cite{Borzi07}, kagome
lattice materials \cite{Yin18}, heavy fermion systems \cite{Okazaki11}, as well as nickel-based
superconductors \cite{Eckberg20}. In particular, in the context of iron-based superconductors
\cite{Chuang10,Gallais13,Massat16}, the experimental observations have shown a striking enhancement
of nematic fluctuations centred at optimal tuning of superconductivity. Moreover, the
nematic-fluctuation-enhanced superconductivity in nickel-based superconductors has been observed
experimentally \cite{Eckberg20}. In a strongly correlated electron system, the strong electron
correlation induces the system to find new way to lower its total energy, often by spontaneous
breaking of the native symmetries of the lattice. These experimental observations
\cite{Chuang10,Gallais13,Massat16,Borzi07,Yin18,Okazaki11,Eckberg20} together with the experimental
detection in cuprate superconductors \cite{Vojta09,Fradkin10,Fernandes19} therefore indicate that
the electronic nematicity is a common phenomenon in strongly correlated electron systems, and then
a characteristic feature in the complicated phase diagram is the interplay between the electronic
nematicity and superconductivity.

\section*{Acknowledgements}

ZC, XM, and SF are supported by the National Key Research and Development Program of China, and the
National Natural Science Foundation of China (NSFC) under Grant Nos. 11974051 and 11734002. HG is
supported by NSFC under Grant Nos. 11774019 and 12074022, and the Fundamental Research Funds for the
Central Universities and HPC resources at Beihang University.



\end{document}